\documentclass[letterpaper, 10pt, conference]{ieeeconf}  % Comment this line out if you need a4paper

\IEEEoverridecommandlockouts                              % This command is only needed if 
\usepackage{graphicx}
\usepackage{amsmath}
\usepackage[psamsfonts]{amssymb}
\usepackage{amsxtra}
\usepackage{threeparttable}
\usepackage{url}
\usepackage{cite}
\usepackage{multirow}

\title{\LARGE \bf
Chromatic and High--frequency cVEP--based BCI Paradigm
}

\author{Daiki Aminaka$^{1}$, Shoji Makino$^{1}$, and Tomasz M. Rutkowski$^{1,2,3,*}$% <-this % stops a space
\thanks{*This research was supported in part by the Strategic Information and Communications R\&D Promotion Program (SCOPE) no. 121803027 of The Ministry of Internal Affairs and Communication in Japan, and by KAKENHI, the Japan Society for the Promotion of Science, grant no.~$24243062.$}% <-this % stops a space
\thanks{$^{1}$Daiki Aminaka, Shoji Makino and Tomasz M. Rutkowski are with Life Science Center of TARA and Department of Computer Science,
        University of Tsukuba, 1-1-1 Tennodai Tsukuba Ibaraki, Japan. {\tt\small tomek@bci-lab.info} \qquad {\tt\small http://bci-lab.info/}}
\thanks{$^{2}$Tomasz M. Rutkowski is also with RIKEN Brain Science Institute, Wako-shi, Japan.}
\thanks{$^{3}$Tomasz M. Rutkowski is the corresponding author.} 
\thanks{$^{*}$Tomasz M. Rutkowski was supported in part by YAMAHA Corporation.}       
        }

\begin{document}

\maketitle
\thispagestyle{empty}
\pagestyle{empty}

%%%%%%%%%%%%%%%%%%%%%%%%%%%%%%%%%%%%%%%%%%%%%%%%%%%%%%%%%%%%%%%%%%%%%%%%%%%%%%%%
\begin{abstract}
We present results of an approach to a code--modulated visual evoked potential (cVEP) based brain--computer interface (BCI) paradigm using four high--frequency flashing stimuli. To generate higher frequency stimulation compared to the state--of--the--art cVEP--based BCIs, we propose to use the light--emitting diodes (LEDs) driven from a small micro--controller board hardware generator designed by our team. The high--frequency and green--blue chromatic flashing stimuli are used in the study in order to minimize a danger of a photosensitive epilepsy (PSE). We compare the the green--blue chromatic cVEP--based BCI accuracies with the conventional white--black flicker based interface.
The high--frequency cVEP responses are identified using a canonical correlation analysis (CCA) method. 
\end{abstract}

%%%%%%%%%%%%%%%%%%%%%%%%%%%%%%%%%%%%%%%%%%%%%%%%%%%%%%%%%%%%%%%%%%%%%%%%%%%%%%%%
\section{INTRODUCTION}

A brain computer interface (BCI) is a technology that utilizes human neurophysiological signals for a direct brainwave--based communication with an external environment, and without depending on any muscle or peripheral nervous system activities~\cite{bciBOOKwolpaw}. Particularly, in the case of patients suffering from locked--in--syndrome (LIS)~\cite{alsTLSdiagnosis1966}, such technology could help them to communicate or complete various daily tasks (type messages on a virtual keyboard or control their environment using a computer, etc). This technology shall create a feasible option for amyotrophic lateral sclerosis (ALS) or coma patients to communicate with their families, friends or caretakers by using their brainwaves only~\cite{bciBOOKwolpaw}. 

In this paper, we report on the visual BCI paradigm, which has been inspired by the three seminal papers in this field, namely a code--modulated visual evoked potential (cVEP) response--based BCI reported in~\cite{bin2011high}; a steady--state visual evoked potential (SSVEP) classification with a canonical correlation analysis (CCA)~\cite{SSVEP-CCA}; and a green--blue flashing--based SSVEP BCI~\cite{sakurada2014use}. The chromatic green--blue cVEP stimulus brings a new option of a lower danger of the a photosensitive epilepsy (PSE)~\cite{PSE-origin,brain2007} seizures comparing to the classical SSVEP based BCI paradigms~\cite{sakurada2014use,Bakardjian201034,daikiAPSIPA2014}. The flashing frequencies in a range of $15\sim25$~Hz are the most provocative for the PSE (the frequencies of $5\sim65$~Hz are still dangerous)~\cite{PSE-origin}.

We also propose to employ the light--emitting diodes (LEDs) for the code modulated visual evoked potential (cVEP) generation with an application of higher flashing frequencies ($40$~Hz), comparing to the classical computer display (with limited refreshing rates allowing for at maximum $30$~Hz stimulus generation) approaches~\cite{bciBOOKwolpaw,cVEP-research,Bakardjian201034}. The higher flashing frequencies allow for an increase of an information transfer rates (ITR), of which low scores were a limitation of the state--of-the--art cVEP--based BCI paradigm~\cite{bin2011high}.
The cVEP is a natural response for the visual stimulus with specific code--modulated sequences~\cite{cVEP-research,bin2011high}. It is generated by the brain when the user gazes at light source which flashes the specific code--modulated sequence. The cVEP--based BCI belongs to the stimulus--driven BCIs, which do not required longer training comparing to the imagery--driven paradigms~\cite{bciBOOKwolpaw}.
In order to further confirm results from the recent studies on the chromatic (green--blue) flashing--based paradigms~\cite{daikiAPSIPA2014,sakurada2014use}, we compare in this paper the proposed BCI paradigm with a classical white--black cVEP modality.

From now on the paper is organized as follows. In the next section we describe materials and methods used in this study. Discussion of the obtained results follows and conclusions, together with future research directions, summarize the paper.

\section{MATERIALS AND METHODS}

The experiments reported in this paper were performed in the Life Science Center of TARA, University of Tsukuba, Japan.
All the details of the experimental procedures and the research targets of the cVEP--based BCI paradigm were explained in detail to the eight users, who agreed voluntarily to participated in the study.
The electroencephalogram (EEG) cVEP--based BCI experiments were conducted in accordance with \emph{The World Medical Association Declaration of Helsinki - Ethical Principles for Medical Research Involving Human Subjects}. 
The experimental procedures were approved and designed in agreement with the ethical committee guidelines of the Faculty of Engineering, Information and Systems at University of Tsukuba, Tsukuba, Japan (experimental permission no.~$2013R7$). 
The average age of the users was of $26.9$ years old (standard deviation of $7.3$ years old; seven males and one female).

\begin{figure}[t]
	\vspace{.2cm}
	\begin{center}
	\includegraphics[width=0.8\linewidth]{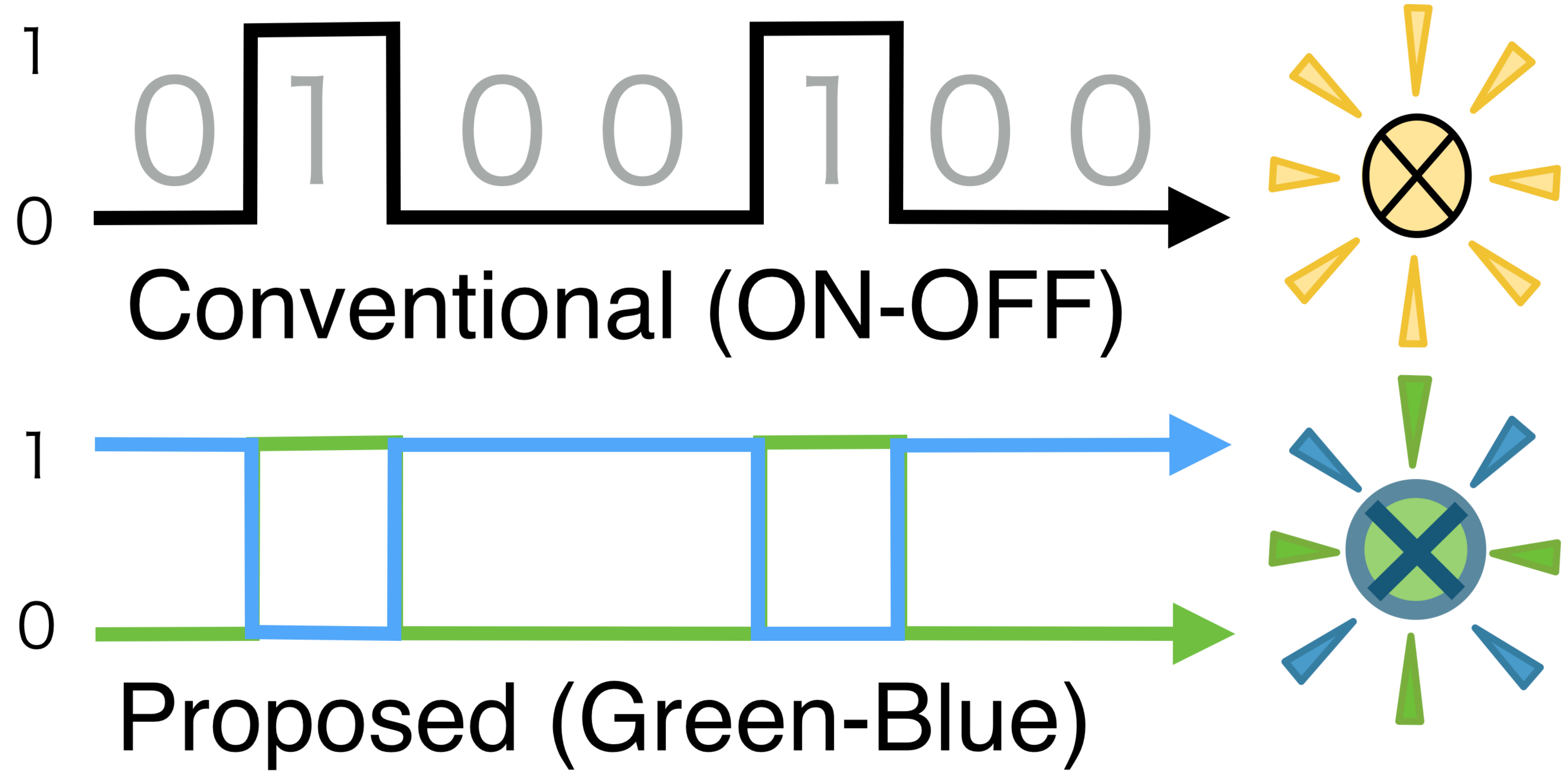}
	\end{center}
	\caption{A diagram explaining a difference in flashing stimuli between the conventional monochromatic (white--black) and the proposed chromatic (green--blue) flickering patterns created with square cVEP--modulated waves generated by the \emph{ARDUINO UNO} micro--controller board in the presented study.}\label{fig:squareWave}
\end{figure}

\begin{figure}[t]
	\vspace{.2cm}
	\begin{center}
	\includegraphics[width=0.9\linewidth]{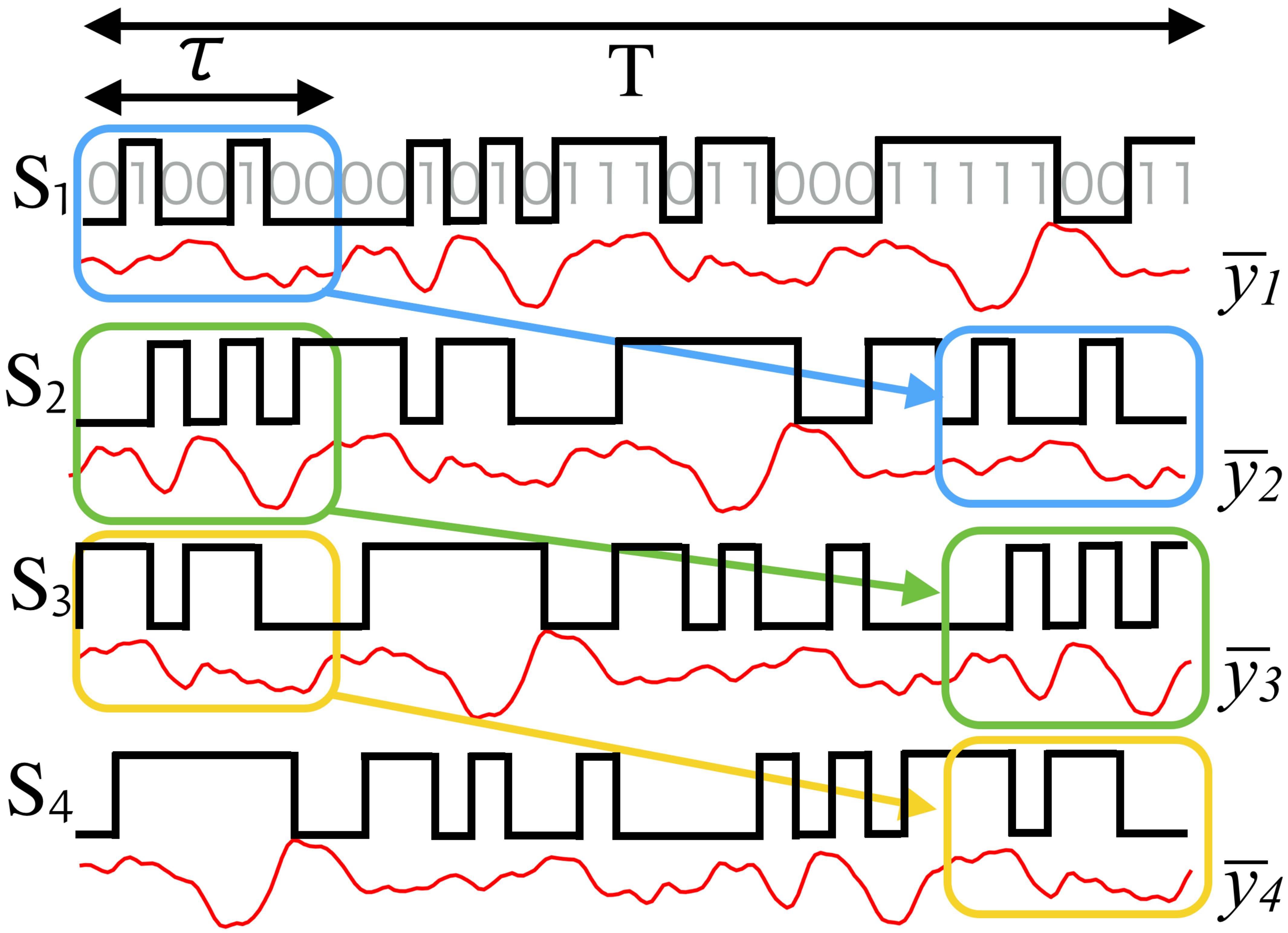}
	\end{center}
	\caption{A diagram illustrating a procedure to construct four $m-sequences$ denoted by ${\bf s}_i, (i=1,2,3,4)$. Each sequence is shifted by $\tau=7$~bits and it has a period $T$. The traces denoted by ${\bf\bar{y}}_j, (j=1,2,3,4)$ are the averaged EEG cVEP responses.}\label{fig:shifted}
\end{figure}

The visual stimuli were flashed via the RGB LEDs as square waves generated by the \emph{ARDUINO UNO} micro--controller board as shown in Figure~\ref{fig:squareWave}. The generator program was written by our team using $C$--language.
In this study we used $m-sequence$ encoded flashing patterns~\cite{bin2011high} to create four commands of the cVEP--based BCI paradigm.
The $m-sequence$ is a binary pseudorandom sequence, which could be generated using the follwoing equation,
\begin{equation}
	x(n) =  x(n-p) \oplus x(n-q), \quad (p > q) \label{eq:mSEQ}
\end{equation}
where $x(n)$ is the $n^{th}$ element of the $m-sequence$ obtained by the exclusive--or (XOR) operation, denoted by $\oplus$, using the two preceding elements indicated by $p$ and $q$. In this project $p = 5$ and $q = 2$ were chosen.
An initial binary sequence was chosen, to create the full $m-sequence$ used in the equation~(\ref{eq:mSEQ}), as follows,
\begin{equation}
	{\mathbf x}_{initial} = [0, 1, 0, 0, 1]. \label{eq:iniSEQ}
\end{equation}
Finally, the $31$ bits long sequence was generated based on the above initial sequence as in equation~(\ref{eq:iniSEQ}) and depicted as a vector ${\bf s}_1$ in Figure~\ref{fig:shifted}.

\begin{figure}[t]
	\vspace{.2cm}
	\begin{center}
	\includegraphics[width=0.7\linewidth]{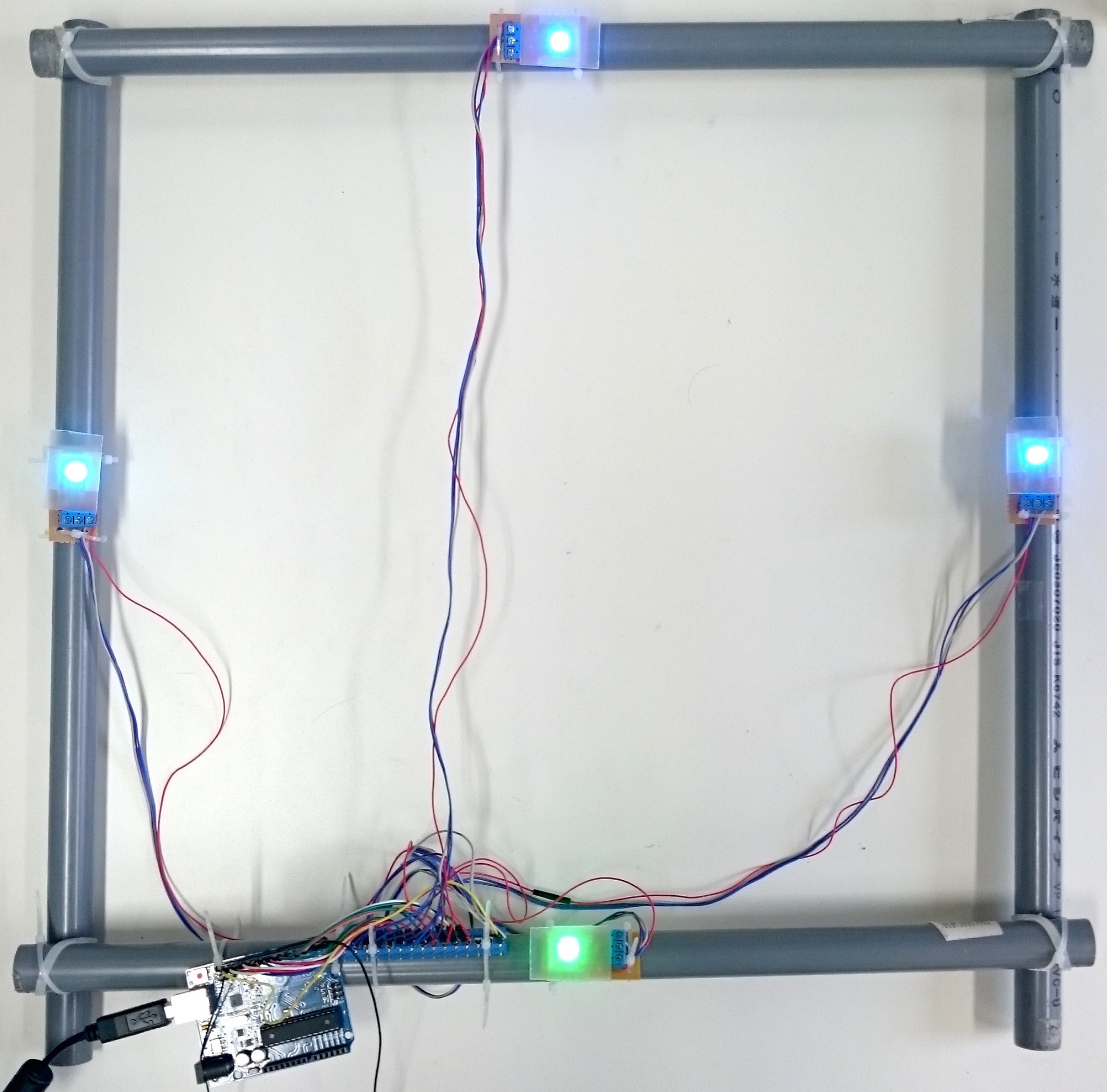}
	\end{center}
	\caption{The visual stimulation device used in the presented study. The top, bottom, right and left LEDs' flickering patterns correspond to ${\bf s}_i, (i=1,2,3,4)$ $m-sequences$, respectively (see the $m-sequence$ shifted patterns Figure~\ref{fig:shifted}). The LEDs were arranged at edge centers of the $45.5\times41.5$~cm rectangular frame as shown in the above photograph.}\label{fig:LEDdisplay}
	\vspace{-.2cm}
\end{figure}

The interesting $m-sequence$ feature, which is very useful for the cVEP--based BCI paradigm design, is an unique autocorrelation function. The autocorrelation function has only a single peak at the period sample value. If the $m-sequence$ period is $N$, the autocorrelation function will result with values equal to $1$ at $0, N, 2N,\ldots$ and $1/N$ otherwise. It is also possible to introduce a circular shift of the $m-sequence$ denoted by $\tau$, to create a set of $m-sequences$ with also shifted autocorrelation functions, respectively.
In this study, the shifted time length has been defined as $\tau = 7$~bits. Three additional sequences has been generated using shifting by $\tau$, $2\cdot\tau$ and $3\cdot\tau$, respectively, as shown also in Figure~\ref{fig:shifted}.
During the online cVEP--based BCI experiments the four LEDs continued to flash simultaneously using the time--shifted $m-sequences$ as explained above. Two $m-sequence$ period lengths have been tested to investigate whether they would affect the cVEP response? The conventional full $m-sequence$ period of $T = 516.7$~ms (based on the conventional computer screen refresh rate of $60$~Hz and referred as ``a low flashing frequency'') and the proposed $T = 387.5$~ms (referred as ``a high flashing frequency'') have been tested. The two monochromatic (white--black) and chromatic (green--blue) LEDs for the proposed cVEP--based BCI settings used in the experiments reported in this paper are presented in Figure~\ref{fig:LEDdisplay}. 
\begin{table}[b]
	\begin{center}
	\caption{EEG experiment condition details}
	\label{tab:EEGsetup}
	\renewcommand{\arraystretch}{1.3}
	\begin{tabular}{| l | l | }
	\hline
	Number of users				& $8$ \\ \hline
	Single session length		& $8$ and $11$~seconds \\ \hline
	$m-sequence$ lengths $T$	& $516.7$~and~$387.5$~ms \\ \hline
	Shifts $\tau$				& $116.7$ and $87.5$~ms  \\ \hline
	\multirow{2}{*}{EEG recording system}							
								& g.USBamp by g.tec with active wet \\
								& (gel--based) g.LADYbird electrodes \\ \hline
	\multirow{2}{*}{EEG electrode locations}		& \textsf{O1, O2, Po3, Po4, P1, P2, Oz} \\
								& and \textsf{Poz}\\ \hline	
	Reference and ground			& Left earlobe and \textsf{FPz} \\ \hline
 	\multirow{2}{*}{Notch filter}				& Butterworth $4^{th}$~order with rejection \\
								&  band of $48 \sim 52$~Hz \\ \hline
	\multirow{2}{*}{Band--pass filter}			& Butterworth $8^{th}$~order with pass  \\
								& band of $5 \sim 100$~Hz \\ \hline
	LED positions					& $45.5\times41.5$~cm rectangle edge centers \\ \hline
	\end{tabular}
	\end{center}
\end{table}
During the cVEP--based BCI EEG experiments the users were seated on a comfortable chair in front of the LEDs (see Figure~\ref{fig:LEDdisplay}). The distance between user's eyes and LEDs was about $30 \sim 50$~cm (chosen by the users for a comfortable view of all LEDs).
An ambient light was moderate as in a typical office.
The EEG signals were captured with a portable EEG amplifier system g.USBamp from g.tec Medical Engineering, Austria. Eight active wet (gel--based) g.LADYbird EEG electrodes were connected to the head locations as in an extended $10/10$ international system~\cite{bciBOOKwolpaw}. These positions were decided due to the visual cortex responses targeting experiment~\cite{bciBOOKwolpaw}. Details of the EEG experimental set up are summarized in Table~\ref{tab:EEGsetup}. The sampling frequency was set to $512$~Hz and a notch $4^{th}$~order Butterworth IIR filter at rejection band of $48\sim52$~Hz was applied to remove power line interference of $50$~Hz. Moreover, the $8^{th}$ order Butterworth IIR band--pass filter at pass band of  $5 \sim 100$~Hz was applied to remove eye blinks and high frequency muscle--originating noise.
The OpenViBE~\cite{OpenViBE} bio--signal data acquisition and processing environment, together with in--house programmed in Python extensions, were applied to realize the online cVEP--based BCI paradigm. To avoid user's eye blinks, each trial to gaze at a single LED was separated with pauses during the experimental sessions (sessions length details in Table~\ref{tab:EEGsetup}). All the participating users were instructed not to move their neck when changing their eye--gaze among the four LEDs to reduce electromyographic noise. No eye--movement interference affected the proposed cVEP BCI.
In the data acquisition phase, first the users gazed at top flashing LED in order to collect classifier training dataset, as instructed verbally by the experimenter conducting the study. 
Twenty $m-sequence$ cycles were repeated in a single EEG capturing session. In short, sixty cVEPs to $m-sequence$ based flashing were collected for each direction in a single experimental trial. The triggers indicating  the onsets of the $m-sequences$ were sent to g.USBamp directly from the \emph{ARDUINO UNO} micro--controlled to mark the beginning of each cVEP response.
Finally, four experiment types were conducted for each user: the conventional low frequency; the proposed high frequency; and in each above setting in the two color modes using white--black and green--blue flashing LEDs.

A canonical correlation analysis (CCA)~\cite{SSVEP-CCA} method  was used in this study to classify the cVEP responses based on the user's intentional choices (gazes).
In the training session a single dataset containing the cVEP responses to top flashing LED was used. The remaining three cVEP responses were constructed by shifting the top LED response by $\tau$, $2\cdot\tau$ and $3\cdot\tau$ (see Figure~\ref{fig:shifted} for stimulus design explanation). 
We applied an analysis of multichannel correlation coefficients among the reference $m-sequences$ and captured cVEPs~\cite{bin2011high}. We used the CCA method to identify the gazed by the user flickering patterns. The cVEP response processing and classification steps were as follows:
\begin{enumerate}
\item Capturing the EEG cVEPs ${\bf{y}}_i, (i=1,2,3,4)$ obtained in response to time shifted $m-sequences$ ${\bf s}_i$ (see Figure~\ref{fig:shifted} for stimulus creation details). 
\item Averaging the captured $j$ cVEPs as $y_{i,j}(t)$ for each target $i$ separately. The averaged responses ${\bf\bar{y}}_i$ were defined as the reference data. In this study, the number of cVEPs was $N = 60$. The averaging procedure was as follows,
\begin{equation}
	{\bf\bar{y}}_i = \frac{1}{N}\sum^{N}_{j=1}{\bf y}_{i,j}.
\end{equation}
\item For the test purposes, $M, (M\ll N),$ cVEPs recorded in separate sessions were used for the method evaluation. In this study $M = 5$. The averaged test datasets were calculated as follows,
\begin{equation}
	{\bf\bar{z}}_{k,l} = \frac{1}{M}\sum^{l+M-1}_{m=l}{\bf z}_{k,m},
\end{equation}
where $k = 1,2,3,4$ represented the stimulus number and  $l= 1, 2, \ldots, N-M+1$ was the test dataset number.
\item In order to identify (classify) the attended stimulus the CCA was calculated for test ${\bf\bar{z}}_{k,l}$ versus the reference ${\bf\bar{y}}_i$ datasets as,
\begin{equation}
	\rho_i = \frac{a^{\mathrm{T}}{R_{\bar{{\bf z}}_{k,l}\bar{{\bf y}}_i}}{b}}{\sqrt{a^{\mathrm{T}}{R_{\bar{{\bf z}}_{k,l}\bar{{\bf z}}_{k,l}}}{a}}\sqrt{b^{\mathrm{T}}{R_{\bar{{\bf y}}_i\bar{{\bf y}}_i}}{b}}},
\end{equation}
where the $R_{\bar{{\bf z}}_{k,l}\bar{{\bf y}}_i}$, $R_{\bar{{\bf z}}_{k,l}\bar{{\bf z}}_{kl}}$, $R_{\bar{{\bf y}}_i\bar{{\bf y}}_i}$ were the respective covariance matrices; $a$ and $b$ were the canonical correlation vectors. The maximum value of correlation coefficient identified by an index $c$  was used to classify the attended stimuli, which pointed the correlated reference as,
\begin{equation}
	\rho_c = \max\{\rho_1, \rho_2, \rho_3, \rho_4\}.
\end{equation}
\end{enumerate}
The results of the above procedure applied to data recorded in the cVEP--based BCI experiments with eight subjects are discussed in the following section.

\section{RESULTS}

\begin{figure*}[t]
	\vspace{.2cm}
	\begin{center}
	\includegraphics[width=0.99\linewidth]{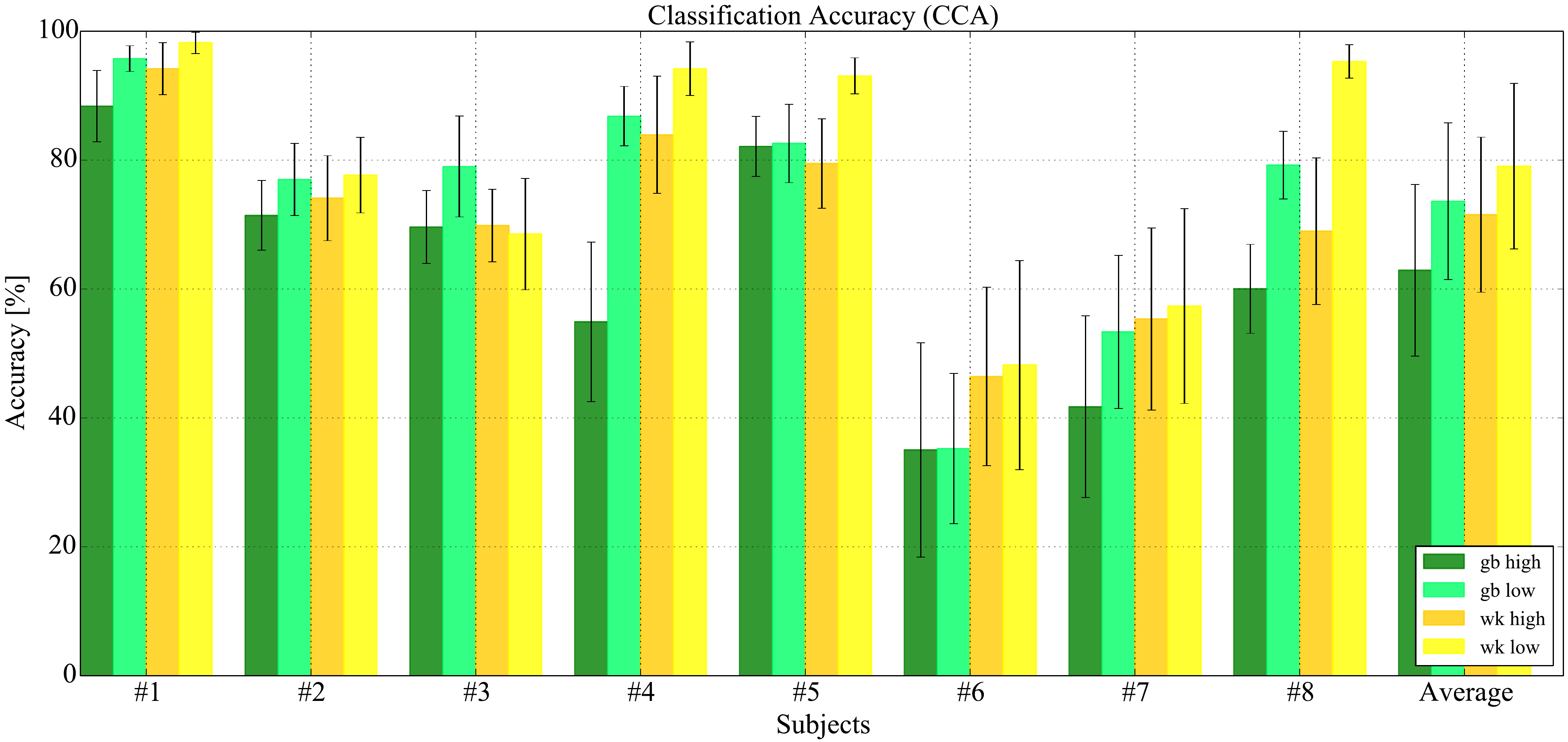}
	\end{center}
	\vspace{-.4cm}
	\caption{The results of CCA--based classification from the eight users participating in the study presented in form of bar plots with standard errors. There are four results depicted for each user, namely from the green--blue high frequency (green); green--blue conventional low frequency (light green); white--black higher frequency (orange); and white--black lower frequency cases (yellow), respectively. The theoretical chance level of the experiments was of $25\%$.}\label{fig:barPlotResult}
\end{figure*}

Results of the conducted cVEP--based BCI paradigm experiments are summarized in form of accuracies as bar plots with standard errors is in Figure~\ref{fig:barPlotResult}.
The accuracies were calculated in experiments using the proposed green--blue high frequency, the green--blue conventional low frequency, the white--black high frequency, and the low frequency white--black flashing settings, respectively. The theoretical chance level of all experiments was of $25\%$.
To investigate flickering frequency effect on cVEP classification (discriminability) accuracies, we applied pairwise Wilcoxon--test for a statistical analysis of median difference significances, because all the accuracy results were not normally distributed. The test was applied for pairs of green--blue flashing in high ($40$~Hz) versus low ($30$~Hz) frequencies and similarly for white--black setting accuracies, respectively. The results of the statistical analyses resulted with significant $p_{gb}<0.029$ for green--blue setting, which supported the project initial hypothesis of the chromatic cVEP modality feasibility. The white--black setting resulted with non--significant differences at a level of $p_{wb} < 0.143$. A comparison for significance of frequencies across to color settings (chromatic versus monochromatic) resulted with non--significant differences for high frequencies with $p_{high}> 0.05$. The low frequencies with significantly different at a level of $p_{low}<0.034$.

\section{CONCLUSIONS}

The proposed LED flashing and cVEP response--based BCI paradigm with the chromatic green--blue stimuli has been discussed in this paper. We tested positively, for the online BCI feasibility, the higher frequency cVEP and CCA classification--based method in comparison to the classical low frequency cVEP stimuli. Also the the chromatic green--blue light effect resulted with encouraging results. 

The conducted experiments to verify the feasibility of the proposed method confirmed successfully our research hypothesis based on the results obtained from healthy eight users. All of the cVEP--based BCI accuracies scored above the theoretical chance levels and there were also $100\%$ accuracies reported.

For the future research, we plan to investigate upper limits of the stimulus frequency and an optimization of the  $m-sequences$ in order to create the even better visual BCI.

\section*{ACKNOWLEDGMENTS}

We would like to thank Dr. Andrzej Cichocki from RIKEN Brain Science Institute, Japan, and Dr. Koich Mori from Research Institute of National Rehabilitation Center for Persons with Disabilities, Japan, for very stimulating discussions about chromatic SSVEP--based BCI which stimulated this project. 

%%%%%%%%%%%%%%%%%%%%%%%%%%%%%%%%%%%%%%%%%%%%%%%%%%%%%%%%%%%%%%%%%%%%%%%%%%%%%%%%

%References are important to the reader; therefore, each citation must be complete and correct. If at all possible, references should be commonly available publications.

\bibliographystyle{IEEEtran}
\bibliography{daiki}

%\begin{thebibliography}{99}
%\bibitem{c1} G. O. Young, 
%\end{thebibliography}

\addtolength{\textheight}{-12cm}  % This command serves to balance the column lengths

\end{document}